\documentstyle [12pt]{article}

\advance \textheight by \topskip
\def\nsection#1{\section{#1}\setcounter{equation}{0}}

\begin{document}

\begin{center}

{ \large \bf   DEFORMED HEISENBERG ALGEBRA,\\[0.2cm] 
FRACTIONAL SPIN FIELDS\\[0.2cm] AND\\[0.2cm]
 SUPERSYMMETRY WITHOUT FERMIONS}\\
\vskip1.0cm
{ \bf
Mikhail S. Plyushchay\footnote{On leave from the
Institute for High Energy Physics,
Protvino, Moscow Region, Russia; E-mail: mikhail@posta.unizar.es}
}\\[0.4cm]
{\it Departamento de F\'{\i}sica Te\'orica, Facultad de Ciencias}\\
{\it Universidad de Zaragoza, Zaragoza 50009, Spain}\\[0.5cm]
{\bf Annals of Physics  245 (1996) 339-360}

\vskip1.0cm

                            {\bf Abstract}
\end{center}
Within a group-theoretical approach to  the description of
(2+1)-dimensional anyons, the minimal covariant set of linear
differential equations is constructed for the fractional spin
fields with the help of the deformed Heisenberg algebra (DHA),
$[a^{-},a^{+}]=1+\nu K$, involving the Klein operator $K$,
$\{K,a^{\pm}\}=0$, $K^{2}=1$.  The connection of the minimal set
of equations with the earlier proposed `universal' vector set of
anyon equations is established.  On the basis of this algebra, a
bosonization of supersymmetric quantum mechanics is carried out.
The construction comprises the cases of exact and spontaneously
broken $N=2$ supersymmetry allowing us to realize a Bose-Fermi
transformation and spin-1/2 representation of SU(2) group in
terms of one bosonic oscillator.  The construction admits an
extension to the case of OSp(2$\vert$2) supersymmetry, and, as a
consequence, both applications of the DHA  turn out to be
related.  A possibility of `superimposing' the two applications
of the DHA for constructing a supersymmetric (2+1)-dimensional
anyon system is discussed.  As a consequential result we point
out that $osp(2|2)$ superalgebra is realizable as an operator
algebra for a quantum mechanical 2-body (nonsupersymmetric)
Calogero model.

\newpage

\nsection{Introduction}

In this work, we consider two related applications of the
deformed Heisenberg algebra, which was introduced by Vasiliev in
the context of higher spin algebras \cite{vas}.  Subsequently,
the generalized and modified (extended) version of this algebra
was used in refs. \cite{bri1}--\cite{bri2} under investigation
of the quantum mechanical N-body Calogero model \cite{caloj},
related to the (1+1)-dimensional anyons \cite{lm1,hlm}.
Moreover, the extended version turned out to be useful in
establishing the links between Knizhnik-Zamolodchikov equations
and Calogero model \cite{bv} (see the paper
\cite{oper} for further references on this model being an
interesting example of one-dimensional quantum integrable
systems).  The applications of the deformed Heisenberg algebra
to be considered here are the construction of the minimal set of
linear differential equations for (2+1)-dimensional fractional
spin fields (anyons) and the bosonization of supersymmetric
quantum mechanics. We also present some speculation on the
possibility to `superimpose' these two applications for
constructing a supersymmetric (2+1)-dimensional anyon system.

The considerable interest to the (2+1)-dimensional anyons, i.e.
particles with fractional spin and statistics \cite{any1}, is
conditioned nowadays by their applications to the theory of
planar physical phenomena: fractional quantum Hall effect and
high-$T_{c}$ superconductivity \cite{appl}.  Anyons also attract
a great attention due to their relationship to the different
theoretical fields of research such as conformal field theories
and braid groups (see, e.g., refs. \cite{fub}--\cite{ler}).

{}From the field-theoretical point of view, such particles can
be described in two, possibly related, ways. The first way
consists in organizing a statistical interaction of the scalar
or fermionic field with the Chern-Simons U(1) gauge field, that
changes spin and statistics of the matter field \cite{sem}. In
this approach, manifestly gauge-invariant nonlocal field
operators, constructed from the initial gauge noninvariant
matter fields following the line integral prescription  of
Schwinger, carry fractional spin and satisfy anyonic permutation
relations \cite{sred}.  Such redefined matter fields are given
on the one-dimensional path ---  unobservable `string' going to
the space infinity and attached to the point in which the
initial matter field is given. Therefore in this approach the
anyonic field operators are path-dependent and multivalued
\cite{sred} (see also ref. \cite{fro1}).  The initial Lagrangian
for the statistically charged matter field can be rewritten in
the decoupled form in terms of the anyonic gauge-invariant
matter fields and Chern-Simons gauge field.  However, such a
formal transition to the free nonlocal anyonic fields within a
path integral approach is accompanied by the appearance of the
complicated Jacobian of the transformation \cite{sreedpr}, and
this means a nontrivial relic of the statistical gauge field in
the theory.  Therefore, the Chern-Simons gauge field approach
does not give a minimal description of (2+1)-dimensional anyons.

Another, less developed way consists in attempting to describe
anyons within the group-theoretical approach analogously to the
case of integer and half-integer spin fields, without using
Chern-Simons U(1) gauge field constructions.  The program of
this approach \cite{ply2}--\cite{pld} consists in constructing
equations for (2+1)-dimensional fractional spin field,
subsequent identifying corresponding field action and, finally,
in realizing a quantization of the theory to reveal a fractional
statistics.  Within this approach, there are, in turn, two
related possibilities: to use many-valued representations of the
(2+1)-dimensional Lorentz group SO(2,1), or to work with the
infinite-dimensional unitary representations of its universal
covering group,  $\overline{\rm SO(2,1)}$ (or $\overline{\rm
SL(2,R)}$, isomorphic to it).  Up to now, it is not clear how to
construct the action functionals corresponding to the equations
for the fractional spin field carrying many-valued
representations of SO(2,1) \cite{ply2,for,ply1}.  On the other
hand, different variants of the equations and some corresponding
field actions were constructed with the use of the unitary
infinite-dimensional representations of $\overline{\rm SL(2,R)}$
\cite{tors}--\cite{pld}.  Nevertheless, the problem of
quantizing the theory is still open here. This is connected with
the search for the most appropriate set of initial equations for
subsequent constructing the action as a basic ingredient for
quantization of the theory. Besides, there is a difficulty in
quantization of such a theory related to the infinite-component
nature of the fractional spin field which is used to describe in
a covariant way one-dimensional irreducible representations of
the (2+1)-dimensional quantum mechanical Poincar\'e group
$\overline{\rm ISO(2,1)}$, specified by the values of mass and
arbitrary (fixed) spin. Due to this fact, an infinite set of the
corresponding Hamiltonian constraints must be present in the
theory to exclude an infinite number of the `auxiliary' field
degrees of freedom and leave in the theory only one physical
field degree of freedom.  This infinite set of constraints
should appropriately be taken into account. But, on the other
hand, the infinite-component nature of the fractional spin field
indicates the possible hidden nonlocal nature of the theory,
and, therefore, can be considered in favour of existence of the
anyonic spin-statistics relation \cite{fro1,fro2} for the
fractional spin fields within the framework of the
group-theoretical approach.  So, calling fractional spin fields
as anyons, we bear in mind such a hypothetical spin-statistics
relation.

Here, we shall consider the problem of constructing the minimal
covariant set of linear differential equations for
(2+1)-dimensional fractional spin fields within a framework of
the group-theoretical approach to anyons, developing the
investigation initiated in refs.  \cite{cor,pld}. We shall see
that the minimal set of linear differential equations plays for
anyons the role analogous to the role played by the Dirac and
Jackiw-Templeton-Schonfield equations \cite{jts} in the cases of
the spinor and topologically massive vector gauge fields,
respectively.  At the same time, these linear differential
equations are similar to the (3+1)-dimensional Dirac
positive-energy relativistic wave equations \cite{dirac}.  As it
has been declared above, the construction will be realized with
the help of the deformed Heisenberg algebra \cite{vas}.  This
algebra involves the so called Klein operator as an essential
object, which introduces $Z_{2}$-grading structure on the Fock
space of the deformed bosonic oscillator.  Such a structure, in
turn, is an essential ingredient of the supersymmetry considered
in ref. \cite{mac} as a hidden supersymmetry of the deformed
bosonic oscillator.  Using this observation, we shall bosonize
the supersymmetric quantum mechanics through the realization of
$N=2$ superalgebra on the Fock space of the deformed (or
ordinary, undeformed) bosonic oscillator.  Moreover, we shall
show  that the two applications of the deformed Heisenberg
algebra turn out to be related through the more general
OSp(2$|$2) superalgebraic structure.

The paper is organized as follows.  In section 2, starting from
the presentation of the main idea of the group-theoretical
approach, we consider a `universal' (but nonminimal) covariant
vector set of linear differential equations for the fractional
spin fields \cite{cor} and formulate the problem of constructing
the minimal spinor set of linear differential equations.  Such a
construction is realized in section 4 with the help of the
deformed Heisenberg algebra, which itself is considered in
section 3. In the latter section we shall show, in particular,
that the Vasiliev deformed bosonic oscillator, given by this
algebra, and the q-deformed Arik-Coon \cite{qac} and
Macfarlane-Biedenharn oscillators \cite{qmb} have a general
structure of the generalized deformed oscillator considered in
ref. \cite{qd}.  Section 5 is devoted to realization of $N=2$
supersymmetric quantum mechanics on the Fock space of one
(ordinary or deformed) bosonic oscillator.  First, we realize
the simplest superoscillator in terms of creation and
annihilation bosonic operators. Such a construction gives us a
possibility to realize a Bose-Fermi transformation in terms of
one bosonic oscillator and construct a spin-1/2 representation
of the SU(2) group on its Fock space. We shall show that the
realization of N=2 supersymmetric quantum mechanics on the Fock
space of the bosonic oscillator is achieved due to a specific
nonlocal character of the supersymmetry generators.  Hence, from
the point of view of a nonlocality, the bosonization scheme
turns out to be similar to the above mentioned Chern-Simons
gauge field constructions for anyons \cite{sem,sred}.  Then, we
shall generalize the constructions to the more complicated $N=2$
supersymmetric systems,  in particular, corresponding to the
Witten supersymmetric quantum mechanics with odd superpotential.
In conclusion of this section, we shall demonstrate that the
construction can be extended to the case of OSp(2$\vert$2)
supersymmetry.  As a consequence, we reveal $osp(2\vert 2)$
superalgebra in the form of the operator superalgebra for the
quantum mechanical 2-body (nonsupersymmetric) Calogero model
\cite{caloj}.

Section 6 is devoted to the discussion of the results, open
problems and possible generalizations of the constructions.

\nsection{`Universal' vector set of equations}

Within a group-theoretical approach to anyons, relativistic
field with fractional (arbitrary) spin $s$ can be described by
the system of the Klein-Gordon equation
\begin{equation}
(P^{2}+m^{2})\Psi=0
\label{kle}
\end{equation}
and linear differential equation
\begin{equation}
(PJ-sm)\Psi=0.
\label{maj}
\end{equation}
Here $P_\mu =-i\partial_\mu$, the metric is
$\eta_{\mu\nu}=diag(-1,1,1)$, and operators $J^{\mu}$,
$\mu=0,1,2,$  being the generators of the group $\overline{\rm
SL(2,R)}$, form the $sl(2)$ algebra
\begin{equation}
[J_{\mu},J_{\nu}]=-i\epsilon_{\mu\nu\lambda}J^{\lambda},
\label{alg}
\end{equation}
where $\epsilon_{\mu\nu\lambda}$ is an antisymmetric tensor
normalized so as $\epsilon^{012}=1$.  We suppose that field
$\Psi=\Psi^{n}(x)$ is transformed according to the one of the
infinite-dimensional unitary irreducible representations (UIRs)
of the group $\overline{\rm SL(2,R)}$:  either of the discrete
type series $D^{\pm}_{\alpha}$, or of the principal or
supplementary continuous series $C^{\theta}_{\sigma}$
\cite{bar}.  These representations are characterized by the
value of the Casimir operator $J^2$ and by eigenvalues $j_0$ of
the operator $J_0$. In the case of the representations
$D^{\pm}_{\alpha}$, we have $J^2 =-\alpha(\alpha-1)$,
$\alpha>0$, and  $j_0 =\pm(\alpha+n)$, $n=0,1,2,\ldots,$ and,
therefore, these representations are half-bounded.  In the case
of the representations of the continuous series
$C^\theta_\sigma$, $J^{2}=\sigma$ and $j_0 =\theta+n$,
$\theta\in [0,1)$, $n=0,\pm 1,\pm 2, \ldots,$ where $\sigma\geq
1/4$ for the principal series and $0<\sigma<1/4$,
$\sigma>\theta(1-\theta)$ for the supplementary series
\cite{bar}.

The system of eqs. (\ref{kle}) and (\ref{maj}) has nontrivial
solutions under the coordinated choice of the representation and
parameter $s$, and as a result, these equations fix the values
of the $\overline{\rm ISO(2,1)}$ Casimir operators, which are
the operators of squared mass,  $M^{2}=-P^{2}$, and spin,
$S=PJ/M$.  In particular, choosing a representation of the
discrete series $D^{\pm}_{\alpha}$ and parameter
$s=\varepsilon\alpha$, $\varepsilon=\pm 1$, we find that these
equations have nontrivial solutions describing the states with
spin $s=\varepsilon\alpha$, mass $M=m$ and energy sign
$\epsilon^0=\pm\varepsilon$.  In this case eq. (\ref{maj})
itself is the (2+1)-dimensional analog of the Majorana equation
\cite{maj}  giving the spectrum of the quantized model of
relativistic particle with torsion
\cite{tors}.  The spectrum of this equation contains an infinite number of
massive states with spin-mass dependence
\begin{equation}
M_{n}=m\frac{\alpha}{|S_{n}|},\quad
S_{n}=\epsilon (\alpha+n),
\label{majsp}
\end{equation}
and, moreover, comprises massless and tachyonic states.
Therefore, from the point of view of the Majorana equation
spectrum, the role of the Klein-Gordon equation consists in
singling out only one state corresponding to $n=0$ from infinite
spectrum (\ref{majsp}) and in getting rid off the massless and
tachyonic states.

The system of equations (\ref{kle}) and (\ref{maj}) has an
essential shortcoming: they are completely independent, unlike,
e.g., the systems of Dirac and Klein-Gordon equations for the
case of spinor field or of Jackiw-Templeton-Schonfeld and
Klein-Gordon equations for a topologically massive vector U(1)
gauge field \cite{jts}. Therefore, they are not very suitable
for constructing the action and quantum theory of the fractional
spin fields, and it is necessary to find out more convenient set
of linear differential equations,  which would be a starting
point for a subsequent realization of the program described
above.

In the recent paper \cite{cor}, the following covariant vector set of
linear differential equations
for a field with  arbitrary fractional spin
has been constructed by the author and J.L. Cort\'es:
\begin{eqnarray}
&V_{\mu}\Psi=0,&
\label{vector}\\
&V_{\mu}=\alpha P_{\mu}-i\epsilon_{\mu\nu\lambda}P^{\nu}J^{\lambda}
+\varepsilon mJ_{\mu},&
\label{lvector}
\end{eqnarray}
where $\varepsilon=\pm 1$, and we suppose that $\alpha$ is an
arbitrary dimensionless parameter.  This set of three equations
has the following remarkable property.  {}From the very beginning
one can suppose only that $J_{\mu}$ are the generators of the
(2+1)-dimensional Lorentz group satisfying commutation relations
(\ref{alg}), not fixing at all the choice of the concrete
(reducible or irreducible) representation. Then, multiplying
eqs. (\ref{vector}) by the operators $J_{\mu}$, $P_{\mu}$ and
$\epsilon_{\mu\nu\lambda}P^{\nu}J^{\lambda}$, we arrive at the
Klein-Gordon equation (\ref{kle}) and Majorana-type equation
(\ref{maj}) with $s=\varepsilon\alpha$ as a consequence of the
initial system of equations (\ref{vector}), and moreover, we
also find that the equation
\begin{equation}
(J^2 +\alpha(\alpha-1))\Psi=0
\label{irr}
\end{equation}
is a consequence of eqs. (\ref{vector}).  Eq. (\ref{irr}) is
nothing else as the condition of irreducibility of corresponding
representation.  When $\alpha>0$, the system of eqs.
(\ref{kle}), (\ref{maj}) and (\ref{irr}) has nontrivial
solutions only for the choice of the unitary
infinite-dimensional representations of the discrete type series
$D^\pm_{\alpha}$, whereas for $\alpha=-j<0$, $j$ being integer
or half-integer, nontrivial solutions take place only under the
choice of the $(2j+1)$-dimensional nonunitary representations
$\tilde{D}{}_j$ of the group $\overline{\rm SL(2,R)}$.  Such
$(2j+1)$-dimensional nonunitary representations are related to
the corresponding unitary representations $D_j$ of the group
$SU(2)$, and, in particular, in the simplest cases $j=1/2$ and
$j=1$ they describe the spinor and vector representations of the
(2+1)-dimensional Lorentz group \cite{cor}.  In these two cases
eqs. (\ref{vector}) can be reduced to the Dirac and to the
Jackiw-Temleton-Schonfeld equations, respectively.  As soon as
the representation is chosen (and, so, eq. (\ref{irr}) is
satisfied identically), the following identity takes place for
the vector differential operatorR$\ $R$V_\mu$:
\begin{eqnarray}
&R^\mu V_\mu \equiv 0,&
\nonumber\\
&R_\mu= \left( (\alpha-1)^2 \eta_{\mu\nu} -
i(\alpha-1)\epsilon_{\mu\nu\lambda} J^{\lambda}
+J_{\nu} J_{\mu} \right)P^\nu.&
\nonumber
\end{eqnarray}
This identity means that after fixing a choice of an admissible
representation, three equations (\ref{vector}) become to be
dependent ones, and in general case any two of them are
independent (in the above mentioned cases of the finite
dimensional nonunitary spinor and vector representations, eqs.
(\ref{vector}) contain only one independent equation).
Therefore, the total set of three equations is necessary only to
have an explicitly covariant set of linear differential
equations.

Thus, eq. (\ref{vector}) is the set of linear differential
equations which itself, unlike the set of initial equations
(\ref{kle}) and (\ref{maj}), fixes the choice of the
representations $D^{\pm}_{\alpha}$ for the description of
fractional spin fields in 2+1 dimensions, and, on the other
hand, establishes some links between the description of ordinary
bosonic integer and fermionic half-integer spin fields, and the
fields with arbitrary spin. At the same time we conclude that
the minimal number of independent linear differential equations
for fractional spin fields is equal to two.  Therefore, the
corresponding minimal covariant set of such equations must have
a spinor form.

To construct such a covariant minimal spinor set of equations,
we turn to the deformed Heisenberg algebra \cite{vas}.  This
algebra permits to realize the generators $J_{\mu}$ for the
representations $D^{\pm}_{\alpha}$ in the form of operators
bilinear in the creation-annihilation deformed oscillator
operators $a^{\pm}$.  On the other hand, as we shall see,
operators $a^{\pm}$ together with operators $J_\mu$ will form
$osp(1\vert 2)$ superalgebra, and, as a consequence, two linear
combinations of $a^\pm$ will form a two-component self-conjugate
object which exactly is a (2+1)-dimensional spinor. It is these
two fundamental properties supplied by the deformed bosonic
oscillator \cite{vas} that will help us to `extract the square
root' from the basic equations (\ref{kle}) and (\ref{maj}), and
to construct a minimal covariant set of linear differential
equations for fractional spin fields.

\nsection{Deformed Heisenberg algebra}

So, let us consider the algebra \cite{vas}
\begin{eqnarray}
&[a^{-},a^{+}]=1+\nu K,&\label{hei}\\
&K^{2}=1,\quad Ka^{\pm}+a^{\pm}K=0,&
\label{klein}
\end{eqnarray}
with the mutually conjugate
operators $a^{-}$ and $a^{+}$, $(a^{-})^{\dagger} =a^+$,
self-conjugate Klein operator $K$
and real deformation parameter $\nu$.
In the undeformed case ($\nu=0$), eq. (\ref{klein}) can be considered
simply as a relation defining the Klein operator $K$.

Under application of algebra (\ref{hei}), (\ref{klein}) to the
quantum mechanical 2-body Calogero model, the Klein operator $K$
is considered as an independent operator being an operator of
the permutation of two identical particles, whereas the
creation-annihilation operators $a^\pm$ are constructed from the
relative coordinate and momentum operators of the particles,
\begin{equation}
a^{\pm}=\frac{q\mp ip}{\sqrt{2}},
\label{apq}
\end{equation}
and in  representation where the operator of the relative coordinate $q$
to be diagonal, the momentum operator is realized in the form
\cite{bri1}--\cite{bri2}:
\begin{equation}
p=-i\left(\frac{d}{dq}-\frac{\nu}{2q}K\right).
\label{pdK}
\end{equation}
Therefore, in such a case algebra (\ref{hei}), (\ref{klein}) is
an extension of the Heisenberg algebra.  On the other hand, one
can realize operator $K$ in terms of the creation-annihilation
operators themselves
\cite{pld,mac},
and in this case algebra (\ref{hei}), (\ref{klein}) gives a
specific `$\nu$-deformed' bosonic oscillator.  We shall understand
algebra (\ref{hei}), (\ref{klein}) in this latter sense,
i.e. as a deformed Heisenberg algebra.

Let us introduce the Fock-type vacuum,
\[
a^{-}|0>=0,\quad
<0|0>=1,
\quad K|0>=\kappa|0>,
\]
where $\kappa=+1$ or $-1$. Without loss of generality,
we put here $\kappa=+1$.
Then we get the action of the operator $a^{+}a^{-}$ on the
states $(a^{+})^{n}|0>$, $n=0,1,\ldots$,
\begin{equation}
a^{+}a^{-}(a^{+})^{n}|0>=[n]_{\nu}(a^{+})^{n}|0>,
\label{a+a-}
\end{equation}
where
\[
[n]_{\nu}=n+\frac{\nu}{2}
\left(1+(-1)^{n+1}\right).
\]
{}From here we conclude that in the case when
\begin{equation}
\nu >-1,
\label{restr}
\end{equation}
the  space of unitary representation of algebra (\ref{hei}),
(\ref{klein}) is given by the complete set of the normalized
vectors
\[
|n>=\frac{1}{\sqrt{[n]_{\nu}!}}(a^{+})^{n}|0>,\quad
<n|n'>=\delta_{nn'},
\]
where
\[
[n]_{\nu}!=\prod_{k=1}^{n}[k]_{\nu}.
\]
In correspondence with eqs. (\ref{klein}),
the Klein operator $K$ separates the complete set of states $|n>$
into even and odd subspaces:
\begin{equation}
K|n>=(-1)^{n}|n>,
\label{z2}
\end{equation}
and, so, introduces $Z_2$-grading structure on the Fock space of
the deformed ($\nu\neq 0$) or ordinary ($\nu=0$) bosonic
oscillator.  Let us introduce the operators $\Pi_{+}$ and
$\Pi_{-}$,
\begin{equation}
\Pi_{\pm}=\frac{1}{2}(1\pm K),
\label{proj}
\end{equation}
which satisfy the equalities
$\Pi_{\pm}^{2}=\Pi_{\pm},$ $\Pi_{+}\Pi_{-}=0,$ $\Pi_{+}+\Pi_{-}=1$, and
are the projector operators on the even and odd subspaces
of the total Fock space, respectively.
Then, from eqs.
(\ref{a+a-}) and (\ref{z2}) we get the equality:
\begin{equation}
a^{+}a^{-}=N+\nu\Pi_{-},
\label{a+a-n}
\end{equation}
where the number operator $N$, by definition, satisfies the commutation
relations
\[
[a^{-},N]=a^{-},\quad
[a^{+},N]=-a^{+},
\]
and equality $N|0>=0$, and, as a consequence,
\[
N|n>=n|n>.
\]
Using eqs. (\ref{a+a-n}) and (\ref{hei}), we get
\[
a^{-}a^{+}=N+1+\nu\Pi_{+},
\]
and, as a result, we arrive at the following expression for the number
operator in terms of the operators $a^{\pm}$:
\begin{equation}
N=\frac{1}{2}\{a^{-},a^{+}\}-\frac{1}{2}(\nu+1).
\label{ndef}
\end{equation}
Therefore,
due to the completeness of the set of basis states $|n>$,
we can realize the Klein operator $K$ in terms of the operators
$a^{\pm}$ by means of equality (\ref{ndef}):
$
K=\exp{i\pi N},
$
or, in the explicitly hermitian form,
\begin{equation}
K=\cos \pi N.
\label{kcos}
\end{equation}
Hence, we have expressed the Klein operator in terms of the
creation and annihilation operators, and, as a result, realized
algebra (\ref{hei}), (\ref{klein}) as a deformation of the
Heisenberg algebra.

To conclude this section, let us make the following remark.  Using eqs.
(\ref{a+a-n}) and (\ref{kcos}), one can write the relation
\begin{equation}
N=a^{+}a^{-}+\frac{\nu}{2}(\cos\pi N -1).
\label{trans}
\end{equation}
This relation is a transcendent equation for the number
operator given as a function
of the normal product $a^+a^-$. As a result, the
defining relations  of the $\nu$-deformed oscillator (\ref{hei}),
(\ref{klein}) can be reduced to the form
\begin{equation}
a^-a^+=g(a^+a^-),
\label{gener}
\end{equation}
where $g(x)=x+f(x)$, and $f(x)=\nu\cos \pi N(x)$ is the function
of $x=a^+a^-$. The generalized deformed oscillator algebra
(\ref{gener}), containing as other particular cases the algebras
of the $q$-deformed Arik-Coon \cite{qac} and
Macfarlane-Biedenharn \cite{qmb} oscillators, was considered by
Dascaloyannis \cite{qd}, and, so, all these deformed bosonic
oscillators have a general structure (see also ref.
\cite{qmmp}).

\nsection{Spinor system of anyon equations}

Let us consider now the following set of three operators $J_{\mu}$,
bilinear in the creation-annihilation operators of the $\nu$-deformed
oscillator:
\begin{equation}
J_{0}=\frac{1}{4}\{a^{+},a^{-}\},\quad
J_{\pm}=J_{1}\pm iJ_{2}=\frac{1}{2}(a^{\pm})^{2}.
\label{jmu}
\end{equation}
These operators satisfy the algebra (\ref{alg}) of the
generators of $\overline{\rm SL(2,R)}$ group for any value of
the deformation parameter $\nu$ given by eq. (\ref{restr}),
whereas the value of the Casimir operator $J_{\mu}J^{\mu}$ is
given here by the relation
\[
J^2=-\hat{\alpha}(\hat{\alpha}-1),
\]
where
\begin{equation}
\hat{\alpha}=\frac{1}{4}(1+\nu K).
\label{hatal}
\end{equation}
So, realization (\ref{jmu}) gives a reducible representation of
the group $\overline{\rm SL(2,R)}$ on the Fock space of the
$\nu$-deformed bosonic oscillator. As we have pointed out, the
total Fock space can be separated into two subspaces, spanned by
even, $|k>_{+}$, and odd, $|k>_{-},$ states being distinguished
by the Klein operator:
\[
K|k>_\pm =\pm |k>_\pm,
\]
\[
|k>_{+}\equiv |2k>,\quad
|k>_{-}\equiv |2k+1>,\quad
k=0,1,\ldots.
\]
These subspaces are
invariant with respect to the action of operators (\ref{jmu}):
\begin{eqnarray}
J_{0}|k>_\pm&=&(\alpha_\pm+k)|k>_\pm,\quad k=0,1,\ldots,
\nonumber\\
J_{-}|0>_\pm&=&0,\quad
J_{\pm}|k>_\pm\propto|k\pm 1>_\pm,\, k=1,\ldots,
\label{inv}
\end{eqnarray}
and Casimir operator takes on them the constant values,
\begin{equation}
J_{\mu}J^{\mu}|k>_\pm=-\alpha_\pm(\alpha_\pm-1)|k>_\pm,
\label{irreo}
\end{equation}
characterized by the parameters
\begin{equation}
\alpha_{+}=\frac{1}{4}(1+\nu)>0,\quad
\alpha_{-}=\alpha_{+}+\frac{1}{2}>\frac{1}{2}.
\label{aeo}
\end{equation}
Relations (\ref{inv}) and (\ref{irreo}) mean that we have
realized the UIRs of the discrete series $D^{+}_{\alpha_{+}}$
and $D^{+}_{\alpha_{-}}$ of the group $\overline{\rm SL(2,R)}$
on the even and odd subspaces of the total Fock space, spanned
by the states $|k>_{+}$ and $|k>_{-}$, respectively.  The UIRs
of the discrete series $D^{-}_{\alpha_{\pm}}$ can be obtained
from realization (\ref{jmu}) by means of the obvious
substitution
\[
J_{0}\rightarrow -J_{0},\quad
J_{\pm}\rightarrow -J_{\mp}.
\]
Further on, for the sake of simplicity we shall consider only
the case of representations $D^{+}_{\alpha}$.

It is necessary to note here, that the described realization of UIRs
$D^{\pm}_{\alpha_{+}}$ and $D^{\pm}_{\alpha_{-}}$
generalizes the well known realization
of the representations $D^{\pm}_{1/4}$ and $D^{\pm}_{3/4}$
in terms of the ordinary bosonic oscillator ($\nu=0$) \cite{per}.

{}For the construction of the minimal
spinor set of linear differential equations sought for,
we introduce the $\gamma$-matrices in the Majorana representation,
\[
(\gamma^{0})_{\alpha}{}^{\beta}=-(\sigma^{2})_{\alpha}{}^{\beta},\quad
(\gamma^{1})_{\alpha}{}^{\beta}=i(\sigma^{1})_{\alpha}{}^{\beta},\quad
(\gamma^{2})_{\alpha}{}^{\beta}=i(\sigma^{3})_{\alpha}{}^{\beta},
\]
which satisfy the relation
$\gamma^{\mu}\gamma^{\nu}=-\eta^{\mu\nu}+i\epsilon^{\mu\nu\lambda}
\gamma_{\lambda}.$
Here $\sigma^{i}$, $i=1,2,3,$ are the Pauli matrices, and
raising and lowering the spinor
indices is realized  by the antisymmetric tensor
$\epsilon_{\alpha\beta}$, $\epsilon_{12}=\epsilon^{12}=1$:
$f_{\alpha}=f^{\beta}\epsilon_{\beta\alpha},$
$f^{\alpha}=\epsilon^{\alpha\beta}f_{\beta}.$

Consider the spinor type operator
\begin{eqnarray}
L_{\alpha}=
\left(\begin{array}{ccc}
q\\
p
\end{array}\right),
\label{lab}
\end{eqnarray}
constructed from the operators $q$ and $p$,
which, in turn,
are defined by the relation of the form of eq. (\ref{apq})
as linear hermitian combinations of $a^\pm$.
In terms of the operators $L_\alpha$, $\alpha=1,2$,
the corresponding part of the deformed Heisenberg algebra
(\ref{hei}) and (\ref{klein}) is presented as
\[
[L_\alpha,L_\beta]=i\epsilon_{\alpha\beta}\cdot(1+\nu K),\quad
\{K,L_\alpha\}=0.
\]
At the same time, these operators satisfy the following anticommutation
relations:
\begin{equation}
\{L_{\alpha},L_{\beta}\}=4i(J_{\mu}\gamma^{\mu})_{\alpha\beta}.
\label{llanti}
\end{equation}
Hence, operator (\ref{lab}) is the `square root'
operator of the $\overline{\rm SL(2,R)}$ generators (\ref{jmu}).
Calculating the commutators of $L_\alpha$ with $J_\mu$
being the generators of the (2+1)-dimensional Lorentz group,
we get the relation:
\begin{equation}
[J_\mu,L_\alpha]=\frac{1}{2}(\gamma_\mu)_\alpha{}^{\beta}L_\beta,
\label{jlcom}
\end{equation}
which means that the introduced operators (\ref{lab}) indeed
form the (2+1)-dimensional spinor.  Moreover, taking into
account (anti)commutation relations (\ref{alg}), (\ref{llanti})
and (\ref{jlcom}), we conclude that the operators $J_{\mu}$ and
$L_\alpha$ form the $osp(1|2)$ superalgebra
\cite{osp} with the Casimir operator
\begin{equation}
C=J^{\mu}J_{\mu}-\frac{i}{8}L^{\alpha}L_{\alpha}=
\frac{1}{16}(1-\nu^2).
\label{ospcas}
\end{equation}
This, in turn, means that we have constructed the generalization of
the well known representation of $osp(1|2)$ superalgebra
with $C=1/16$, realized by the ordinary bosonic oscillator
($\nu=0$) \cite{next,macmaj},
to the case of $C<1/16$ ($\nu\neq 0$)
(see also ref. \cite{mac}).

Having the spinor object $L_\alpha$, which is a `square root'
operator of $J_\mu$, we can construct a covariant spinor linear
differential operator with dimensionality of mass of the
following most general form not containing a dependence on the
Klein operator $K$:
\begin{equation}
S_{\alpha}=L^{\beta}\left((P\gamma)_{\beta\alpha}+\varepsilon
m\epsilon_{\beta\alpha}\right),
\label{da}
\end{equation}
where, again, $P_{\mu}=-i\partial_{\mu}$ and $\varepsilon=\pm1$.
Therefore, one can consider the spinor set of linear
differential equations
\begin{equation}
S_{\alpha}\Psi=0.
\label{baseq}
\end{equation}
Here we suppose that the field $\Psi=\Psi^{n}(x)$
is an infinite-component function given on the Fock space
of the $\nu$-deformed bosonic oscillator,
i.e. eq. (\ref{baseq}) is a symbolic presentation of the
set of two infinite-component equations
$S_{\alpha}^{nn'}\Psi^{n'}=0$ with
$S_\alpha^{nn'}=<n|S_{\alpha}|n'>$.

Operator (\ref{da}) satisfies the relation
\[
S^{\alpha}S_{\alpha}=L^{\alpha}L_{\alpha}(P^{2}+m^{2}).
\]
Since
$L^{\alpha}L_{\alpha}=-i(1+\nu K)\neq 0$  due to restriction
(\ref{restr}), we conclude that the Klein-Gordon equation
(\ref{kle}) is the consequence of eqs. (\ref{baseq}).
Moreover, we have the relation
\[
L^{\alpha}S_{\alpha}=-4i\left(PJ-\varepsilon m\hat{\alpha}\right),
\]
where $\hat{\alpha}$ is given by eq. (\ref{hatal}).
Decomposing the field $\Psi$ as
$\Psi=\Psi_{+}+\Psi_{-}$,
\begin{equation}
\Psi_{\pm}=\Pi_{\pm}\Psi,
\label{psipm}
\end{equation}
and, so, $\Psi_{+}=\Psi^{2k}$, $\Psi_{-}=\Psi^{2k+1}$,
$k=0,1,\ldots$, we conclude that as a consequence of basic
eqs. (\ref{baseq}), the fields $\Psi_{\pm}$
satisfy, respectively, the equations
\begin{equation}
(PJ-\varepsilon\alpha_{+}m)\Psi_{+}=0,
\label{maje}
\end{equation}
\begin{equation}
\left(PJ-\varepsilon
\left(\alpha_{-}-\frac{1}{2}(1+\nu)\right)m\right)\Psi_{-}=0,
\label{majo}
\end{equation}
where $\alpha_+$ and $\alpha_-$ are given by eq. (\ref{aeo}).
Note, that eq.  (\ref{maje}) is exactly the (2+1)-dimensional
analogue of the Majorana equation.  Taking into account eq.
(\ref{kle}), and passing over to the rest frame ${\bf P}={\bf
0}$ in the momentum representation, we find with the help of eq.
(\ref{jmu}) that eq.  (\ref{maje}) has the solution of the form
$\Psi_{+}\propto \delta(P^{0}-\varepsilon m) \delta({\bf P})
\Psi^{0}$, whereas eq. (\ref{majo}) has no nontrivial solution.
Hence, the pair of equations (\ref{baseq}) has nontrivial
solutions only in the case $\Psi=\Psi_{+}$, describing the field
with spin $s=\varepsilon\alpha_{+}$ and mass $m$.  At $\nu=0$,
eqs. (\ref{baseq}) turn into Volkov-Sorokin-Tkach equations for
a field with spin $s=1/4\cdot \varepsilon$ \cite{stv,sv} being
(2+1)-dimensional analogues of the Dirac (3+1)-dimensional
positive-energy relativistic wave equations \cite{dirac}.

Thus, we have constructed the minimal covariant system of linear
differential equations (\ref{baseq}) for the field with
arbitrary fractional spin, whose value is defined by  the
deformation parameter: $s=\varepsilon\cdot\frac{1}{4}(1+\nu)\neq
0$.

There is the following connection between the spinor operator
(\ref{da}) and the vector operator (\ref{lvector}), which is
valid on the even subspace of the Fock space of the
$\nu$-deformed bosonic oscillator:
\begin{equation}
(\gamma_{\mu})^{\alpha\beta}L_{\alpha}S_{\beta}\Psi_+=V_{\mu}\Psi_+,
\label{spvec}
\end{equation}
where we suppose that $V_\mu$ is given by eq. (\ref{lvector})
with the operators $J_\mu$ being realized in the form
(\ref{jmu}).  {}From here we conclude that the field $\Psi_+$
satisfying spinor system of independent eqs.  (\ref{baseq}),
satisfies also the vector system of dependent equations
(\ref{vector}), and, therefore, the constructed system of
equations (\ref{baseq}) is a fundamental system of linear
differential equations for the fractional spin fields.  We shall
return to the discussion of these equations and connection
(\ref{spvec}) in last section.

\nsection{Bosonization of supersymmetric quantum mechanics}

As it was pointed out in the original papers \cite{vas}, the
deformed Heisenberg algebra (\ref{hei}) and (\ref{klein}) leads
to different higher spin superalgebras. We have shown in the
previous section that the $\nu$-deformed bosonic oscillator
permits to generalize the well known representation of the
$osp(1|2)$ superalgebra, realized by the ordinary bosonic
oscillator ($\nu=0$) and characterized by the value of the
Casimir operator $C=1/16$, to the representations with
$C=\frac{1}{16}(1-\nu^2)$.  Further, we have seen that the Klein
operator introduces $Z_2$-grading structure on the Fock space of
the ordinary or $\nu$-deformed bosonic oscillator.  Such a
structure is an essential ingredient of supersymmetry.
Therefore, we arrive at the natural question: whether it is
possible to use this $Z_2$-grading structure for realizing
representations of $N=2$ supersymmetry in terms of
$\nu$-deformed  or ordinary ($\nu=0$) bosonic oscillator, and,
therefore, for bosonizing supersymmetric quantum mechanics.

In the present section we shall investigate the problem of
bosonization of supersymmetric quantum mechanics \cite{nic,wit}
in the systematic way.  We shall consider the $\nu$-deformed
bosonic oscillator, but it is necessary to stress that every
time one can put $\nu=0$, and therefore,  all the constructions
will also be valid for the case of the ordinary bosonic
oscillator.

We begin with the construction of the mutually conjugate
nilpotent supercharge operators $Q^+$ and $Q^-$,
$Q^+=(Q^-)^\dagger$, in the simplest possible form, linear in
the bosonic operators $a^\pm$, but also containing a dependence
on the Klein operator $K$:
\[
Q^{+}=\frac{1}{2}a^{+}(\alpha+\beta K)+\frac{1}{2}a^{-}(\gamma+\delta K).
\]
The nilpotency condition $Q^{\pm 2}=0$ leads to the restriction on the
complex number coefficients: $\beta=\epsilon\alpha$,
$\delta=\epsilon\gamma$.
Therefore, we find that there are two possibilities for choosing operator
$Q^+$:
\[
Q^{+}_{\epsilon}=(\alpha a^{+}+\gamma a^{-})\Pi_\epsilon,
\quad \epsilon=\pm,
\]
where $\Pi_\pm$ are the projector operators (\ref{proj}).
The anticommutator
$\{Q^{+}_{\epsilon},Q^{-}_{\epsilon}\}$
has here the form:
\[
\{Q^{+}_{\epsilon},Q^{-}_{\epsilon}\}=
a^{+2}\alpha\gamma^{*}+a^{-2}\alpha^{*}\gamma
+\frac{1}{2}\{a^{+},a^{-}\}(\gamma\gamma^{*}+\alpha\alpha^{*})
-\frac{1}{2}\epsilon K
[a^{-},a^{+}](\gamma\gamma^{*}-\alpha\alpha^{*}).
\]
Whence we conclude that if we choose the parameters in such a
way that $\alpha\gamma^{*}=0$, the anticommutator will commute
with the number operator $N$, and, as a consequence, the spectra
of the corresponding Hamiltonians, $H_\epsilon
=\{Q^+_\epsilon,Q^-_\epsilon\}$, $\epsilon=\pm$, will have the
simplest possible form.  Let us put $\alpha=0$ and normalize the
second parameter as $\gamma=e^{i\varphi}$. Since we can remove
this phase factor by the unitary transformation of the operators
$a^{\pm}$, we arrive at the nilpotent operators in the very
compact form:
\begin{equation}
Q^{+}_{\epsilon}=a^{-}\Pi_{\epsilon},\quad
Q^{-}_{\epsilon}=a^{+}\Pi_{-\epsilon}.
\label{qexp}
\end{equation}
They together with the operator
\begin{equation}
H_{\epsilon}=\frac{1}{2}\{a^{+},a^{-}\}
-\frac{1}{2}\epsilon K [a^{-},a^{+}]
\label{hexp1}
\end{equation}
form the $N=2$ superalgebra, which we shall also denote, according to
ref. \cite{next}, as $s(2)$:
\begin{equation}
Q^{\pm2}_{\epsilon}=0,\quad \{Q^{+}_{\epsilon},Q^{-}_{\epsilon}\}
=H_{\epsilon},\quad
[Q^{\pm}_{\epsilon},H_{\epsilon}]=0.
\label{n2s2}
\end{equation}
Note that the hermitian supercharge operators
$Q^{1,2}_{\epsilon}$,
\[
Q^{\pm}_{\epsilon}=\frac{1}{2}(Q^{1}_{\epsilon}\pm iQ^{2}_{\epsilon}),\quad
\{Q^{i}_{\epsilon},Q^{j}_{\epsilon}\}=2\delta^{ij}H_{\epsilon},
\]
have the following compact form:
\begin{equation}
Q^{1}_{\epsilon}=\frac{1}{\sqrt{2}}(q+i\epsilon pK),\quad
Q^{2}_{\epsilon}=\frac{1}{\sqrt{2}}(p-i\epsilon qK)=
-i\epsilon Q^{1}_{\epsilon}K
\label{q12}
\end{equation}
in terms of the coordinate $q$ and momentum
$p$ operators of the deformed bosonic oscillator
introduced by eq. (\ref{apq}).

Consider the spectrum of the supersymmetric Hamiltonian
(\ref{hexp1}). Using the expression for the number operator given
by eq. (\ref{ndef}), we find that
in the case when $\epsilon=-$,
the states $|n>$ are the eigenstates  of the operator $H_{-}$ with
the eigenvalues
\[
E_{n}^{-}=2[n/2]+1+\nu,
\]
where $[n/2]$ means the integer part of $n/2$. Therefore, for
$\epsilon=-$ we have the case of spontaneously broken
supersymmetry, with $E_{n}^{-}>0$ for all $n$ due to restriction
(\ref{restr}).  All the states $|n>$ and $|n+1>$, $n=2k$,
$k=0,1,\ldots$, are paired here in supermultiplets.  For
$\epsilon=+$, we have the case of exact supersymmetry
characterized by the spectrum
\[
E_{n}^{+}=2[(n+1)/2],
\]
i.e. here the vacuum state with $E_{0}^{+}=0$ is a supersymmetry
singlet, whereas $E_{n}^{+}=E_{n+1}^{+}>0$ for  $n=2k+1,
k=0,1,\ldots$.  Hence, we have demonstrated that one can realize
both cases of the spontaneously broken and exact $N=2$
supersymmetries with the help of only one bosonic oscillator,
and in the former case  the scale of supersymmetry breaking is
defined by the deformation parameter $\nu$.

Due to the property $E_{n}^{-}>0$ taking place for $\epsilon=-$,
we can construct the Fermi oscillator operators:
\begin{eqnarray}
f^{\pm}&=&\frac{Q^{\mp}_{-}}{\sqrt{H_{-}}}
\label{fdef}\\
&=&a^{\pm}\cdot \frac{\Pi_{\pm}}{\sqrt{N+\Pi_{+}}},
\nonumber
\end{eqnarray}
 \[
\{f^{+},f^{-}\}=1,\quad
f^{\pm 2}=0,
\]
i.e. one can realize a Bose-Fermi transformation in terms of one
bosonic oscillator.  Let us note that though operators $a^{\pm}$
do not commute with $f^{\pm}$,
\[
[a^{\pm},f^{\pm}]\neq 0,
\]
nevertheless, the operator $H_{\epsilon}$ can be written in the
form of the simplest supersymmetric Hamiltonian of the
superoscillator
\cite{nic}:
\[
H_{\epsilon}=\frac{1}{2}\{a^{+},a^{-}\}+\epsilon \frac{1}{2}[f^{+},f^{-}].
\]

Having  fermionic oscillator variables (\ref{fdef}),
we can construct the hermitian operators:
\[
S_{1}=\frac{1}{2}(f^{+}+f^{-}),\quad
S_{2}=-\frac{i}{2}(f^{+}-f^{-}),\quad
S_{3}=f^{+}f^{-}-1/2.
\]
They act in an irreducible way on every 2-dimensional subspace of states
$(|2k>,|2k+1>)$, $k=0,1,\ldots$, and form $su(2)$ algebra:
\[
[S_{i},S_{j}]=i\epsilon_{ijk}S_{k}.
\]
Due to the relation $S_{i}S_{i}=3/4,$ this means that we have
realized spin-$1/2$ representation of the $su(2)$ algebra in
terms of one bosonic oscillator in contrast to the well known
Schwinger realization of Lie algebra $su(2)$ in terms of two
bosonic oscillators \cite{schw}.  Therefore, one can realize
unitary spin-1/2 representation of SU(2) group on the Fock space
of one bosonic oscillator.

Before passing over to the generalization of the constructions
to the case of more complicated quantum mechanical
supersymmetric systems, corresponding to the systems with
boson-fermion interaction \cite{wit,salhol}, let us make the
following remark on the realized bosonization scheme.

We have pointed out in the first section that the `anyonization'
scheme involving the Chern-Simons gauge field has an essentially
nonlocal nature. Our constructions include the Klein operator as
a fundamental object, which itself is a nonlocal operator, and,
as a result, all the bosonization scheme presented above has an
essentially nonlocal character.  Indeed, in the simplest case of
the ordinary bosonic oscillator ($\nu=0)$, in the Schr${\rm
\ddot{o}}$dinger representation the Klein operator (\ref{kcos})
is presented in the nonlocal form
\[
K=\sin(\pi H_{0})
\]
through the Hamiltonian of the linear harmonic oscillator
\[
H_{0}= \frac{1}{2}\left(-\frac{d^{2}}{dq^{2}}+q^{2}\right).
\]
As a consequence, the supersymmetric Hamiltonian (\ref{hexp1})
and hermitian supercharge operators (\ref{q12}) also have a
nonlocal form:
\begin{equation}
H_{\epsilon}=H_{0}-\frac{1}{2}\epsilon\sin(\pi H_{0}),
\label{hnon}
\end{equation}
\begin{equation}
Q^1_\epsilon =\frac{1}{\sqrt{2}}
\left(q+\epsilon\frac{d}{dq} \sin(\pi H_{0})\right),\quad
Q^2_\epsilon
=-\frac{i}{\sqrt{2}} \left(\frac{d}{dq}+\epsilon q\sin(\pi H_{0})\right),
\label{qnon}
\end{equation}
and, therefore, in this sense, our bosonization constructions
turn out to be similar to the Chern-Simons gauge field
constructions for (2+1)-dimensional anyons \cite{sem,sred}.

Let us show how the constructions can be generalized to the case
corresponding to the more complicated quantum mechanical supersymmetric
systems \cite{wit,salhol,gen}.
To this end, consider the operators
\begin{equation}
\tilde{Q}^{\pm}_{\epsilon}=A^{\mp}\Pi_{\pm\epsilon}
\label{tilq}
\end{equation}
with odd mutually conjugate operators $A^{\pm}=A^{\pm}(a^{+},a^{-})$,
$A^{-}=(A^{+})^{\dagger}$, $KA^{\pm}=-A^{\pm}K$.
These properties of $A^{\pm}$ guarantee that the
operators
$\tilde{Q}^{\pm}_{\epsilon}$ are, in turn, mutually conjugate,
$\tilde{Q}^{-}_{\epsilon}=
(\tilde{Q}^{+}_{\epsilon})^{\dagger}$,
and nilpotent:
\begin{equation}
(\tilde{Q}{}^{\pm}_{\epsilon})^{2}=0.
\label{tqq}
\end{equation}
One can take the anticommutator
\begin{eqnarray}
\tilde{H}_{\epsilon}&=&
\{\tilde{Q}^{+}_{\epsilon},\tilde{Q}^{-}_{\epsilon}\}
\label{the}\\
&=&\frac{1}{2}\{A^{+},A^{-}\}-\frac{1}{2}\epsilon K
[A^{-},A^{+}].
\label{htilde}
\end{eqnarray}
as the Hamiltonian,
and get the  $s(2)$ superalgebra given by relations
(\ref{tqq}), (\ref{the}) and by the commutator
\begin{equation}
[\tilde{H}_{\epsilon},\tilde{Q}^{\pm}_{\epsilon}]=0.
\label{thq}
\end{equation}
Such a generalized construction given by eqs. (\ref{tilq}) and
(\ref{htilde}) corresponds to $N=2$ supersymmetric systems
realized in ref. \cite{gen} on the Fock space of independent
bosonic and fermionic oscillators.  In particular, choosing the
operators $A^{\pm}$ as
\[
A^{\pm}=\frac{1}{\sqrt{2}}(\mp ip+W(q))
\]
with $W(-q)=-W(q)$, in the case of Heisenberg algebra ($\nu=0$)
we get for supersymmetric Hamiltonian (\ref{htilde}) the form
\[
\tilde{H}_{\epsilon}=\frac{1}{2}\left(-\frac{d^2}{dq^2}+W^{2}-\epsilon K
\frac{dW}{dq}\right)
\]
corresponding to the Witten supersymmetric quantum mechanics \cite{wit}
with odd superpotential $W$.

Concluding the bosonization constructions, we note that the case
of the S(2) supersymmetry is contained in the more broad
OSp(2$|$2) supersymmetry, whose superalgebra, as well as
$osp(1|2)$ superalgebra considered in the previous section,
contains $sl(2)$ algebra (\ref{alg}) as a subalgebra \cite{osp}.
Taking into account this observation, let us show that this more
broad OSp(2$|$2) supersymmetry also can be bosonized.  Indeed,
one can check that the even operators
\begin{equation}
T_3=2J_{0},\quad  T_\pm=J_\pm,\quad
J=-\frac{1}{2}\epsilon K[a^-,a^+],
\label{eveng}
\end{equation}
together with the odd operators
\begin{equation}
Q^\pm =Q^\mp_\epsilon ,\quad
S^\pm=Q^\mp_{-\epsilon},
\label{oddg}
\end{equation}
form the $osp(2|2)$ superalgebra
given by the nontrivial (anti)commutators
\begin{eqnarray}
&[T_3 ,T_\pm ]=\pm 2T_\pm,\quad
[T_- ,T_+ ]=T_3,&\nonumber\\
&\{S^+ ,Q^+ \}=T_+ ,\quad
\{Q^+ ,Q^- \}=T_3 +J,\quad
\{S^+ ,S^- \}=T_3 -J,&\nonumber\\
&[T_+ ,Q^- ]=-S^+ ,\quad
[T_+ , S^- ]=-Q^+,\quad
[T_3 ,Q^+ ]=Q^+,&\nonumber\\
&[T_3, S^- ]=-S^- ,\quad
[J,S^- ]=-S^- ,\quad
[J,Q^+ ]=-Q^+ ,&
\label{osp}
\end{eqnarray}
and corresponding other nontrivial (anti)commutators which can
be obtained from eqs. (\ref{osp}) by hermitian conjugation.
Therefore, even operators (\ref{eveng}) form a subalgebra
$sl(2)\times u(1)$, whereas $s(2)$ superalgebra (\ref{n2s2}), as
a subalgebra, is given by the sets of generators $Q^\pm$ and
$T_3+J$, or $S^\pm$ and $T_3-J$. Hence, the both cases of the
exact and spontaneously broken $N=2$ supersymmetry are contained
in the extended supersymmetry Osp(2$|$2). In eqs. (\ref{eveng})
and (\ref{oddg}) we suppose that operators $J_{\mu}$ and
$Q^\pm_\epsilon$ are realized by  $\nu$-deformed oscillator
operators $a^\pm$ in the form of eqs.  (\ref{jmu}) and
(\ref{qexp}), respectively, whereas the Klein operator $K$ is
realized with the help of eqs. (\ref{kcos}) and (\ref{ndef}).

Superalgebra (\ref{osp}), obviously, also takes place in the
case discussed at the beginning of section 3, when
algebra (\ref{hei}), (\ref{klein}) is considered as an extended
Heisenberg algebra with the Klein operator being the permutation
operator, independent from $a^\pm$, as it happens under
consideration of the quantum mechanical Calogero model
\cite{bri1}--\cite{bri2}.  Therefore, we conclude that
$osp(2|2)$ superalgebra (as well as smaller $s(2)$ superalgebra)
can be realized as an operator superalgebra for the 2-body
(nonsupersymmetric) Calogero model.  Note here that the
supersymmetric extension of the $N$-body Calogero model,
possessing OSp(2$|$2) supersymmetry as a dynamical symmetry, was
constructed by Freedman and Mende \cite{fm}.  The generators of
this dynamical symmetry of the supersymmetric Calogero  model
\cite{fm} were realized by Brink, Hansson and Vasiliev in terms
of bilinears of the modified bosonic creation and annihilation
operators (of the form given by eqs. (\ref{apq}) and (\ref{pdK})
in the case of 2-body model) and of independent fermionic
operators, and  our notation for the generators in eqs.
(\ref{eveng}) and (\ref{oddg}) have been chosen in
correspondence with that in ref.
\cite{bri2}.

\nsection{Discussion and concluding remarks}

We have constructed the minimal covariant set of linear
differential equations (\ref{baseq}) for the fractional spin
fields, which is related to the vector set of equations
(\ref{vector}) via eq. (\ref{spvec}).  The latter set of
equations possesses a property of `universality': it describes
ordinary integer or half-integer spin fields under the choice of
the $(2j+1)$-dimensional nonunitary representations of the group
$\overline{\rm SL(2,R)}$ instead of the infinite-dimensional
unitary representations $D^\pm_\alpha$, necessary for the
description of the fractional spin.  Therefore, it would be
interesting to investigate the minimal spinor set of equations
from the point of view of the possible analogous properties of
universality.  We hope to consider this problem elsewhere.

The minimal set of equations (\ref{baseq}) represents by itself
two independent infinite sets of equations for one
infinite-component field.  Therefore, it is necessary to
introduce some auxiliary fields for the construction of the
corresponding field action and subsequent quantization of the
theory. Investigation of the previous problem as well as the
search for a possible hidden relation of the theory to the
approach involving Chern-Simons U(1) gauge field constructions
\cite{sem,sred} could be helpful for constructing a `minimal'
field action leading to eqs. (\ref{baseq}) and comprising the
minimal number of auxiliary fields. Since both sets of linear
differential equations (\ref{baseq}) and (\ref{vector}), as well
as two other known sets of linear differential equations
considered in refs.
\cite{jac,ply4}, use the half-bounded infinite dimensional
representations of the $\overline{\rm SL(2,R)}$ for the
description of fractional spin fields, one can consider this
fact as an indication on the possible hidden connection of a
group-theoretical approach with the approach involving
Chern-Simons gauge field constructions.  Indeed, the
half-bounded nature of these representations ($n=0,1,2,\ldots)$
could be associated with the half-infinite nonobservable
`string' of the nonlocal anyonic field operators within the
framework of the latter approach.  This is, of course, so far 
pure speculation.

As it has been pointed out in section 3, under application of
algebra (\ref{hei}), (\ref{klein}) to the quantum mechanical
2-body Calogero model, the Klein operator is considered as the
operator of permutation for identical particles on the line,
being independent from the operators $a^\pm$ in the sense of its
realization. In this case algebra (\ref{hei}), (\ref{klein}) is
the extension of the Heisenberg algebra, and operators $a^\pm$
can be realized with the help of relation (\ref{apq}) through
the relative coordinate and momentum operators of the particles,
$q=q_1-q_2$, $p=p_1-p_2$: in representation with the operator
$q$ to be diagonal, the momentum operator is given by eq.
(\ref{pdK}).  Obviously, our constructions from section 4 will
also be valid under consideration of algebra (\ref{hei}),
(\ref{klein}) as the extended one. In this case the field $\Psi$
will depend on two arguments, $\Psi=\Psi(x,q)$, and nontrivial
solutions of eqs.  (\ref{baseq}) will be described by even
fields $\Psi(x,q)=\Psi(x,-q)$.  On the other hand, we have
pointed out in section 1 on the relationship of the Calogero
model to the systems of (1+1)-dimensional anyons \cite{lm1,hlm}.
In particular, as it was shown by Hansson, Leinaas and Myrheim
\cite{hlm}, the system of 2-dimensional anyons in the lowest
Landau level is effectively the system of 1-dimensional anyons,
described by means of the Calogero model.  Therefore, the
described possibility for the reinterpretation of our
constructions in terms of extended Heisenberg algebra means that
the fields $\Psi(x,q)$ can be understood as the fractional spin
fields in (2+1)-dimensional space time, and, at the same time,
as the fields describing the system of two 1-dimensional anyons,
related to the 2-body Calogero model.  Such a reinterpretation
seems to be very attractive from the point of view of possible
revealing spin-statistics relation for (2+1)-dimensional
fractional spin fields within a framework of the approach under
consideration.

We have shown that $osp(2|2)$ superalgebra can be revealed in
the form of an operator (spectrum generating ) algebra for the
2-body (nonsupersymmetric) Calogero model.  It seems that with
the help of generalization of algebra (\ref{hei}), (\ref{klein})
given in refs.
\cite{bri1}--\cite{bri2}, one could also reveal this
superalgebra in the form of the spectrum generating algebra for
the general case of N-body Calogero model.  Note that here we
have an analogy with the case of the ordinary
(nonsupersymmetric) harmonic oscillator, for which the
superalgebra $osp(1|2)$ is the spectrum generating superalgebra
\cite{next}.

As a further generalization of the bosonization constructions
given in section 5, one could investigate a possibility to
bosonize $N>2$ supersymmetric \cite{next} and parasupersymmetric
\cite{rub} quantum mechanical systems.  For the former case, one
could try to use the generalizations of the Klein operator of
the form: $\tilde{K}{}^{l}=1$, $l>2$.  Another interesting
problem is the construction of the classical Lagrangians, which
would lead after quantization to the supersymmetric systems
described in previous section, in particular, to the simplest
system described by Hamiltonian (\ref{hnon}).  The knowledge of
the form of such Lagrangians and corresponding actions
((0+1)-dimensional from the field-theoretical point of view)
could be helpful in possible generalizing the constructions to
the case of the quantum field systems, $(1+1)$-dimensional in
the simplest case. In connection with such hypothetical possible
generalization it is necessary to point out that earlier some
different problem was investigated by Aratyn and Damgaard
\cite{arad}.  They started from the (1+1)-dimensional
supersymmetric field systems, bosonized them with the help of
the Mandelstam nonlocal constructions \cite{mand}, and,  as a
result, arrived at the pure bosonic quantum field systems,
described by the local action functionals.  On the other hand,
an essentially nonlocal form of just the simplest Hamiltonian
(\ref{hnon}) is an indication that the corresponding
generalization of the constructions presented here to the case
of quantum field theory would lead to the nonlocal bosonic
quantum field systems.

To conclude, let us point out the possibility for a
`superposition' of the two applications of the deformed
Heisenberg algebra, presented here.

The essential object for the construction  of the minimal set of
linear differential equations for fractional spin fields is the
$osp(1|2)$ superalgebra realized in terms of the $\nu$-deformed
bosonic oscillator.  This superalgebra is the superalgebra of
automorphisms of the fundamental (anti)commutation relations for
the system containing one bosonic (described by mutually
conjugate coordinate and momentum operators) and one fermionic
(given by the selfconjugate generator of the Clifford algebra
$C_1$) degrees of freedom.  At the same time, it is a subalgebra
of the superalgebra $osp(2|2)$ being a superalgebra of
automorphisms for the system with one bosonic and two fermionic
(given by the two selfconjugate generators of the Clifford
algebra $C_2$) degrees of freedom \cite{next}.  So, the both our
constructions turn out to be related through the $osp(2|2)$
superalgebra (\ref{osp}).  Note, that the operators $a^\pm$ (or
their hermitian linear combinations $L_\alpha$), being the odd
generators of $osp(1|2)$, are presented in terms of odd
generators (\ref{oddg}) of $osp(2|2)$ in the form:
\[
a^\pm =Q^\pm + S^\pm.
\]
The even subalgebra of $osp(1|2)$ is $sl(2)$ algebra with
generators (\ref{jmu}). We have seen that the deformed (or
extended) Heisenberg algebra permits to realize irreducible
representations $D^\pm_{\alpha_{+}}$ and $D^\pm_{\alpha_{-}}$ of
the corresponding group $\overline{\rm SL(2,R)}$, characterized
by the parameters (\ref{aeo}), shifted in 1/2 and contained in
one irreducible representation of $osp(1|2)$ superalgebra,
characterized, in turn, by the Casimir operator (\ref{ospcas}).
The constructed system of eqs.  (\ref{baseq}) has nontrivial
solutions only on the even subspace of the total Fock space,
described by functions $\Psi_+$ (or on even functions
$\Psi(x,q)=\Psi(x,-q)$ in the case of extended Heisenberg
algebra), i.e. only on the space of the $D^\pm_{\alpha_{+}}$
series of representations.  But if we could `enliven' the second
series of representation through the construction of the system
of equations having solutions for both series, we would have the
system of two states with spins shifted in one-half.  If,
moreover, these two states will have equal masses, we would get
a (2+1)-dimensional supersymmetric system of fractional spin
fields.  So, it would be very interesting to investigate the
possibility to supersymmetrize the system of eqs. (\ref{baseq})
by using odd operators (\ref{oddg}) (being spinor operators from
the point of view of $sl(2)$ generators (\ref{jmu})) instead of
the standard introduction of additional Grassmann spinor
variables
\cite{sv}, and therefore, to realize the superposition of the
both constructions presented here.  The work in this direction
is in progress.

\vskip0.3cm
{\bf Acknowledgements}

I thank  J. Ambjorn, P.H. Damgaard, T.H.
Hansson, J.M.  Leinaas, U. Lindstr${\rm \ddot{o}}$m, J.~Myrheim,
A. Niemi, K. Olaussen, V.V.  Sreedhar, P.Di Vecchia and L.C.R.
Wijewardhana for valuable discussions.  I am grateful to Niels
Bohr Institute and Universities of Oslo, Trondheim, Stockholm
and Uppsala, where a part of this work was carried out, for
their kind hospitality, and to NORDITA for a financial support.
This work was also supported in part by MEC-DGICYT, Spain.

\end{document}